\documentclass[%
 preprint,
showpacs,preprintnumbers,
 amsmath,amssymb,
 aps,
prc,
]{revtex4-1}

\usepackage{graphicx}% Include figure files
\usepackage{dcolumn}% Align table columns on decimal point
\usepackage{bm}% bold math
\begin{document}

%\preprint{APS/123-QED}

\title{Jet quenching effects on the direct, elliptic, and triangular flow at RHIC}% Force line breaks with \\

\author{R.~P.~G.~Andrade,$^1$ J.~Noronha,$^1$}
\author{Gabriel S. Denicol$^2$}
\affiliation{$^1$Instituto de F\'{\i}sica, Universidade de S\~{a}o Paulo, C.P. 66318,05315-970 S\~{a}o Paulo, SP, Brazil\\
$^2$Department of Physics, McGill University, 3600 University Street, Montreal, Quebec, H3A, 2T8, Canada}

\date{\today}% It is always \today, today,

\begin{abstract}
In this paper we investigate how the energy and momentum deposited by partonic dijets in the quark-gluon plasma may affect the direct, elliptic and triangular flow of low (and intermediate) $p_T$ hadrons in central Au+Au collisions at RHIC. The dijets are modeled as external sources in the energy-momentum conservation equations for hydrodynamics, which are solved on an event-by-event basis within the ideal fluid approximation. We focus our investigation at  mid-rapidity and solve the hydrodynamic equations imposing boost invariance. Differential anisotropic flow coefficients for $p_T \gtrsim 1$ GeV are found to be significantly enhanced if the dijets deposit on average more than 12 GeV in the QGP (or more than 6 GeV per jet). Because this jet-induced extra anisotropic flow is not related to the fluctuations of the initial geometry of the collision, the correlation between the $v_2$ and $v_3$ coefficients and their corresponding eccentricities is considerably weakened. In addition, we argue that the extra amount of direct flow induced by dijets may be quantified by comparing the azimuthal dependence of dihadron correlations in dijet events with the corresponding quantity obtained in events without dijets. This comparison could be used to give a rough estimate of the magnitude of the effective coupling between the jets and the medium.
\end{abstract}

%\pacs{25.75.-q,12.38.Mh,24.10.Nz,25.75.Ld,25.75.Gz}% PACS, the Physics and Astronomy
\maketitle

\section{\label{sec:introduction} Introduction}

The observation of jet quenching in ultrarelativistic heavy ion collisions performed at the Relativistic Heavy Ion Collider (RHIC) and the Large Hadron Collider (LHC) \cite{Adams:2003kv,Adcox:2001jp,Adams:2003im,Adler:2003ii,CMS:2012aa,Abelev:2012hxa,Aamodt:2010jd} is one of the main evidences that such experiments are in fact able to produce a novel state of bulk nuclear matter, usually called quark-gluon plasma (QGP). Studying and predicting the energy loss of jets formed in such collisions has become one of the most important tasks in the field of heavy ion collisions, serving to further develop our knowledge about the properties of the QGP. On the other hand, while the modification of jets due to the interaction with the medium has been widely investigated (see, for instance, \cite{Burke:2013yra} and references therein), the modification of the medium due to the interaction with jets has not yet been fully explored.

The aim of this article is to motivate a discussion about the effects of dijets in the hydrodynamic evolution of the QGP. Through a simplified model defined on an event-by-event basis, we try to understand such effects in terms of the anisotropic flow parameters $\left\{v_n,\Psi_n \right\}$, the first one being the $n$th Fourier coefficient of the azimuthal distribution of hadrons and the second the respective phase. Depending on the amount of energy-momentum deposited in the medium by the dijets, which depends on how opaque to jets the QGP is, and how fast the energy and momentum deposited in the medium are thermalized, the jet quenching effect can create additional anisotropic flow that is independent on the initial geometry of the matter created in heavy ion collisions.

In our simplified scenario, the amount of energy and momentum deposited in the medium depends basically on three factors: the coupling between the jets and the QGP, the amount of matter along the trajectory of each jet, and the hydrodynamic evolution of the medium. In this approach, we study two different aspects related to the effects of jets in the medium. First, we allow the jets to lose more and more energy and momentum in the medium and try to see how much is necessary to disconnect the flow parameters $\left\{v_n,\Psi_n \right\}$, as a function of $p_T$, from the initial geometric parameters $\left\{\varepsilon_{m,n},\Phi_{m,n} \right\}$, the $n$th eccentricity and the corresponding phase (for a discussion of how the flow parameters are related to the geometry of the initial conditions for hydrodynamics, see for instance Refs. \cite{Alver:2006wh,Alver:2010gr,Alver:2010dn,Qin:2010pf,Bhalerao:2011bp,Bhalerao:2011yg,Teaney:2010vd,Qiu:2011iv,Gardim:2011xv,
Teaney:2012ke,Qiu:2012uy,Niemi:2012aj,Gardim:2012im,Heinz:2013bua,Andrade:2013poa}). Secondly, we discuss how the dihadron angular correlation function may be used to estimate the magnitude of the jet energy loss within the hydrodynamic evolution of the QGP. All the results presented in this article correspond to Au+Au collisions at 200A GeV in the $(0-5)\%$ centrality window.

Now, let us describe how this paper is organized. In Section \ref{sec:model} we give some details about our hydrodynamic model including the equation of state and the decoupling mechanism. In the next section we show that the energy-momentum deposited in the medium by the dijets can be parametrized through a source term in the hydrodynamic equations. In Section \ref{sec:ebe_pro} we discuss the event-by-event procedure used in this article including the modeling of the fluctuating initial conditions. We  then present our results in Section \ref{sec:results} and finish in \ref{sec:conclusion} with our conclusions and outlook. We use hyperbolic coordinates, i.e., $\tau=\sqrt{t^2-z^2}$, $\eta=0.5\ln \left[ \left(t+z \right)/\left(t-z \right)  \right]$ and $\vec{r}=\left(x,y \right)$. In addition, $\hbar=k_{B}=c=1$.

\section{\label{sec:model} Hydrodynamic Model}

Initially, we assume that the energy-momentum deposited in the medium by a partonic jet quickly thermalizes in the QGP \cite{Chaudhuri:2005vc}. Given this hypothesis, the effects of a dijet on the hydrodynamic evolution of the medium can be taken into account through a source term in the energy-momentum conservation equations.  In the ideal fluid approximation, used in this paper, one finds that

\begin{equation}
D_{\mu}T^{\mu \nu}=J^{\nu},
\label{eq:em_cons}
\end{equation}

\noindent
where $T^{\mu \nu}=\omega u^{\mu} u^{\nu} - pg^{\mu \nu}$ is the ideal fluid energy-momentum tensor, $\omega$ is the enthalpy, and $p$ the pressure. The flow 4-velocity $u^\mu$ obeys the normalization $u_\mu u^\mu=1$, $D_\mu$ is the covariant derivative, and $g^{\mu \nu}$ is the metric tensor. The source $J^\nu$ is the 4-current density that describes the dijets (in Section \ref{sec:source} we discuss how this source term may be parametrized).

In our model, Eq.\ (\ref{eq:em_cons}) is solved assuming boost-invariance, i.e., the thermodynamic quantities and the transverse fluid velocity are independent on the rapidity $\eta$ (in addition, the 4-velocity component $u_{\eta}=0$). This approximation was also employed in \cite{Chaudhuri:2005vc}. In the boost-invariant Ansatz the focus is on the transverse expansion around the mid-rapidity. In addition, we assume that the baryon chemical potential and the initial transverse velocity are zero. We note, however, that the dynamics of jets is not boost invariant and a more complete study in full 3+1 hydrodynamics remains to be done. We intend to perform such a study in the near future.

To solve the energy-momentum conservation equations with these assumptions, we apply the relativistic version of the so called Smoothed Particles Hydrodynamics (SPH) approach originally developed in Ref. \cite{Aguiar:2000hw}, which is a suitable tool to deal with irregular distributions of matter. The smoothing SPH parameter is set to be $h=0.3$ fm, which allows for relatively quick computation times while still preserving the important structures present in the initial conditions (for a discussion about the numerical parameters of the 2+1 SPH approach, see Ref. \cite{Andrade:2013poa}). To test the accuracy of our numerical computation, in Ref. \cite{Andrade:2013poa} we showed that our 2+1 ideal hydro code perfectly matches the exact solution for 2+1 ideal hydrodynamics obtained in Ref. \cite{Gubser:2010ze} (see also \cite{Marrochio:2013wla} for the corresponding viscous solution), also known as Gubser flow (in this solution the profile of the hydrodynamic quantities is smooth and $J^{\nu}=0$). We use the equation of state EOS S95n-v1 \cite{Huovinen:2009yb}, which combines results from lattice QCD at high temperatures and the hadron resonance gas equation at low temperatures.

To compute the particle spectrum, we use the Cooper-Frye prescription \cite{Cooper:1974mv}. In this method, the particles escape from the fluid after crossing a hyper-surface of constant temperature, usually called freeze-out temperature, $T_{fo}$ (in Ref. \cite{Andrade:2013poa} we discuss some details about the Cooper-Frye prescription in the 2+1 SPH approach). Because the aim of this paper is to understand the effects of the dijets on the hydrodynamic evolution of the QGP rather than finding the optimal parameters of the model that describe the data, the role of the freeze-out temperature here is only to control the expansion time of the fluid. By using $T_{fo}=0.14$ GeV, which is a typical value in the literature (see, for instance, Ref. \cite{Luzum:2010ae}), the total expansion time in the $\left(0-5 \right)\%$ centrality window is around 15 fm/c, which gives enough time to induce the hydrodynamics effects on the final spectrum of hadrons.

All the results presented in this paper correspond to direct positively charged pions. The implementation of the routines that compute the decay of resonances is in progress. Despite the fact that we cannot make a rigorous comparison between our results and the data, the study of the distribution of pions, directly emitted from the freeze-out hyper-surface, already brings relevant information about the hydrodynamic evolution of the QGP \cite{Andrade:2013poa,Noronha-Hostler:2013gga,Noronha-Hostler:2013ria}.

\section{\label{sec:source} Modeling the source term}

Taking into account that our hydrodynamic model assumes boost invariance, we have to consider the partonic jets traveling at mid-rapidity. For the sake of simplicity, we assume that each parton, which gives rise to a jet, moves at the speed of light. Thus, its trajectory can be parametrized in the following way

\begin{equation}
\vec{r}^{\hspace{0.5mm}\operatorname{jet}}_n \left(\tau \right) = \vec{r}^{\hspace{0.5mm}\operatorname{jet}}_{0n} + \left(\tau - \tau_0 \right) \vec{v}^{\hspace{0.5mm}\operatorname{jet}}_n,
\label{eq:parton_traj}
\end{equation}

\noindent
where $\tau_0=1$ fm is the initial time at which we begin the hydrodynamic evolution and the motion of the partons through the medium. The vectors $\vec{r}^{\hspace{0.5mm}\operatorname{jet}}_n$ and $\vec{v}^{\hspace{0.5mm}\operatorname{jet}}_n$ (with $\left|\vec{v}^{\hspace{0.5mm}\operatorname{jet}}_n \right|=1$) are the position and velocity of the $n$th parton that moves in a straight line on transverse plane at mid-rapidity.

In the scenario described above, the 4-current density $J^{\nu}$, which describes the energy and momentum deposited in the medium by the partonic jets, is light-like. Then, in the laboratory frame, it can be parametrized as (see, for instance, Refs.\ \cite{Chaudhuri:2005vc,Betz:2010qh,Betz:2009su,Torrieri:2009mv,Betz:2008ka,Betz:2008wy})

\begin{equation}
J^{\nu}\left(\tau, \vec{r}\right)= \sum_{n=1}^{n_p} \left(\frac{dE}{d l}\right)_n F \left(\vec{r} - \vec{r}^{\hspace{0.5mm}\operatorname{jet}}_n \left(\tau \right),\tau \right) \left(1,\vec{v}^{\hspace{0.5mm}\operatorname{jet}}_n,0 \right)
\label{eq:current1}
\end{equation}

\noindent
where

\begin{equation}
F \left(\vec{r} - \vec{r}^{\hspace{0.5mm}\operatorname{jet}}_n \left(\tau \right),\tau \right)=\frac{\tau^{-1}}{2 \pi \sigma^2} \exp \left[-\left( \vec{r} - \vec{r}^{\hspace{0.5mm}\operatorname{jet}}_n \left(\tau \right) \right)^2/2\sigma^2 \right],
\label{eq:current2}
\end{equation}

\noindent
$n_p$ is the number of partons and $\left(dE/d l\right)_n$ is the energy loss rate of the $n$th parton with respect to the transverse distance $l=\tau - \tau_0$. In this article, we consider only one dijet per event, which means that $n_p=2$,  $\vec{r}^{\hspace{0.5mm}\operatorname{jet}}_{01}=\vec{r}^{\hspace{0.5mm}\operatorname{jet}}_{02}$ and $\vec{v}^{\hspace{0.5mm}\operatorname{jet}}_1=-\vec{v}^{\hspace{0.5mm}\operatorname{jet}}_2$. The function $F$ describes a source of finite width $\sigma$. To ensure the stability of the numerical computation, this width should be greater than the smoothing SPH parameter $h$ (0.3fm). We set $\sigma=0.6$ fm. Moreover

\begin{equation}
\int F \left(\vec{r} - \vec{r}^{\hspace{0.5mm}\operatorname{jet}}_n \left(\tau \right),\tau \right) \tau dxdyd \eta = \Delta \eta,
\label{eq:norma1}
\end{equation}

\noindent
where the normalization $\Delta \eta$ arises from our assumption regarding boost invariance. In this approximation for the mid-rapidity region, observe that we are not studying the effects on the medium caused by a ``bullet" (possibly creating a Mach cone) which would demand a full 3-dimensional computation such as Ref. \cite{Tachibana:2014lja}. Rather, this model sees the effects on the medium caused by a ``knife" \cite{Chaudhuri:2005vc}, which is parallel to the beam direction and cuts transversely the fluid. As remarked by Ref. \cite{Chaudhuri:2005vc}, one expects that these assumptions regarding the source dynamics provide an upper limit for the effects of jets on the hydrodynamical evolutions of the QGP.

Finally, the energy loss rate can be parametrized as

\begin{equation}
\left(\frac{dE}{d l}\right)_n = \frac{s \left(\vec{r}^{\hspace{0.5mm}\operatorname{jet}}_n \left(\tau \right) \right)}{s_0} \left. \frac{dE}{d l} \right|_0
\label{eq:energy_lost}
\end{equation}

\noindent
where $s \left(\vec{r}^{\hspace{0.5mm}\operatorname{jet}}_n \right)$ is the entropy density computed at the position of the $n$th parton. The parameters $s_0$ and $\left. dE/d l \right|_0$ are the reference entropy density and the reference energy loss rate. In this article $s_0=70$ fm$^{-3}$, which corresponds to the maximum of the average entropy density distribution in the $(0-5)\%$ centrality window. To investigate the effects of the jets on the observables, we vary $\left. dE/d l \right|_0$ from 5 to 20 GeV/fm. The bigger this parameter is, the more energy-momentum the jets deposit in the medium. As we are going to discuss in Section \ref{sec:results}, even for a large energy loss rate, $\left. dE/d l \right|_0\sim$ 20 GeV/fm, the amount of energy deposited in the QGP may be relatively small (approximately 8 GeV per jet, on average). This is a consequence of the hydrodynamic expansion that quickly enhances the mean free path of the partonic jets.

It is important to mention that a more realistic model of jet energy loss can be implemented in our setup and we intend to do that in a future study. However, considering that the purpose of this article is to understand, through the Fourier coefficients of the flow, the fate of the energy and momentum deposited by dijets in the medium, we shall investigate here only the consequences of this simplified energy loss model.

\section{\label{sec:ebe_pro} Event-by-Event Procedure}

\begin{figure*}[ht]
\includegraphics[scale=0.33]{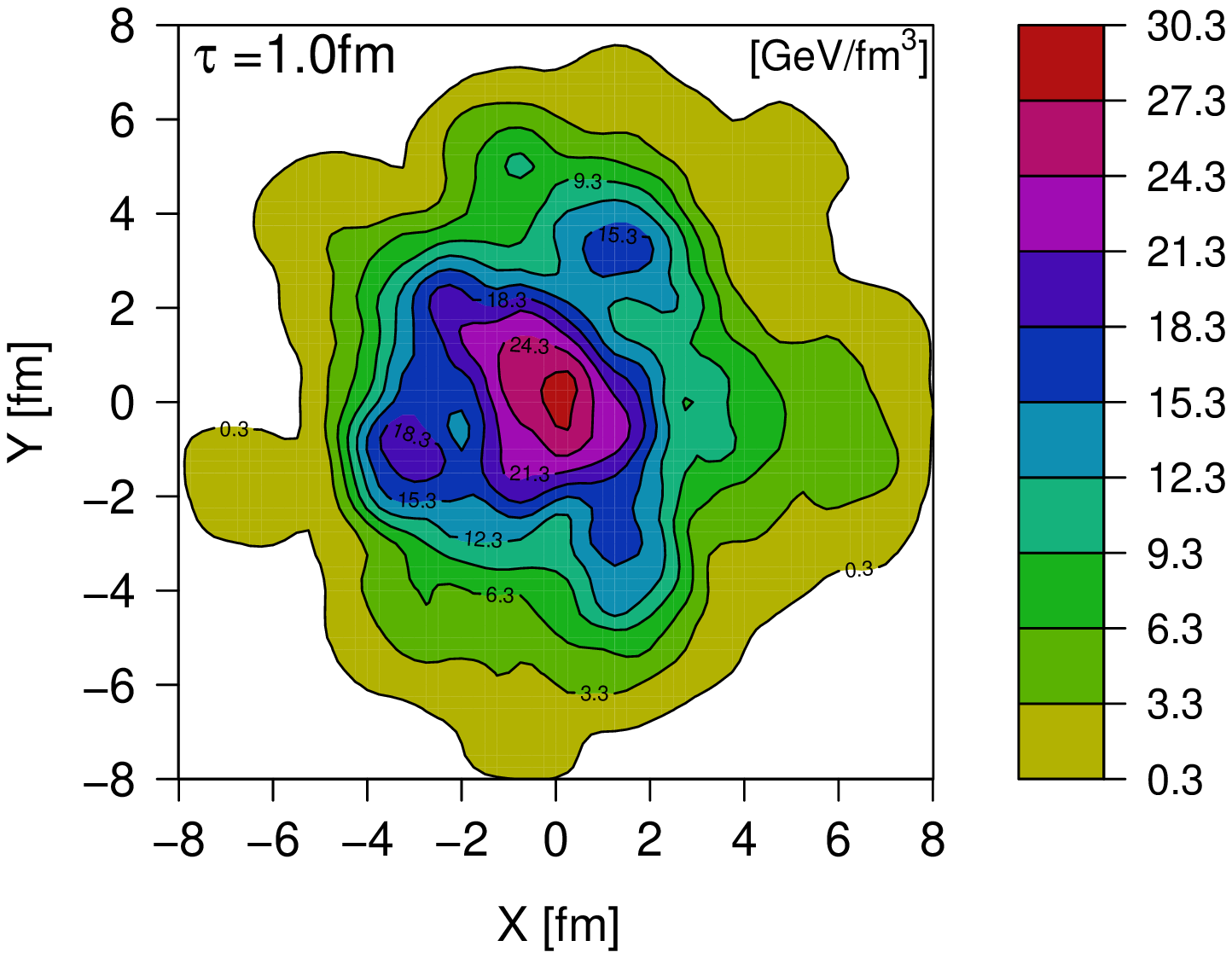}% Here is how to import EPS art
\includegraphics[scale=0.33]{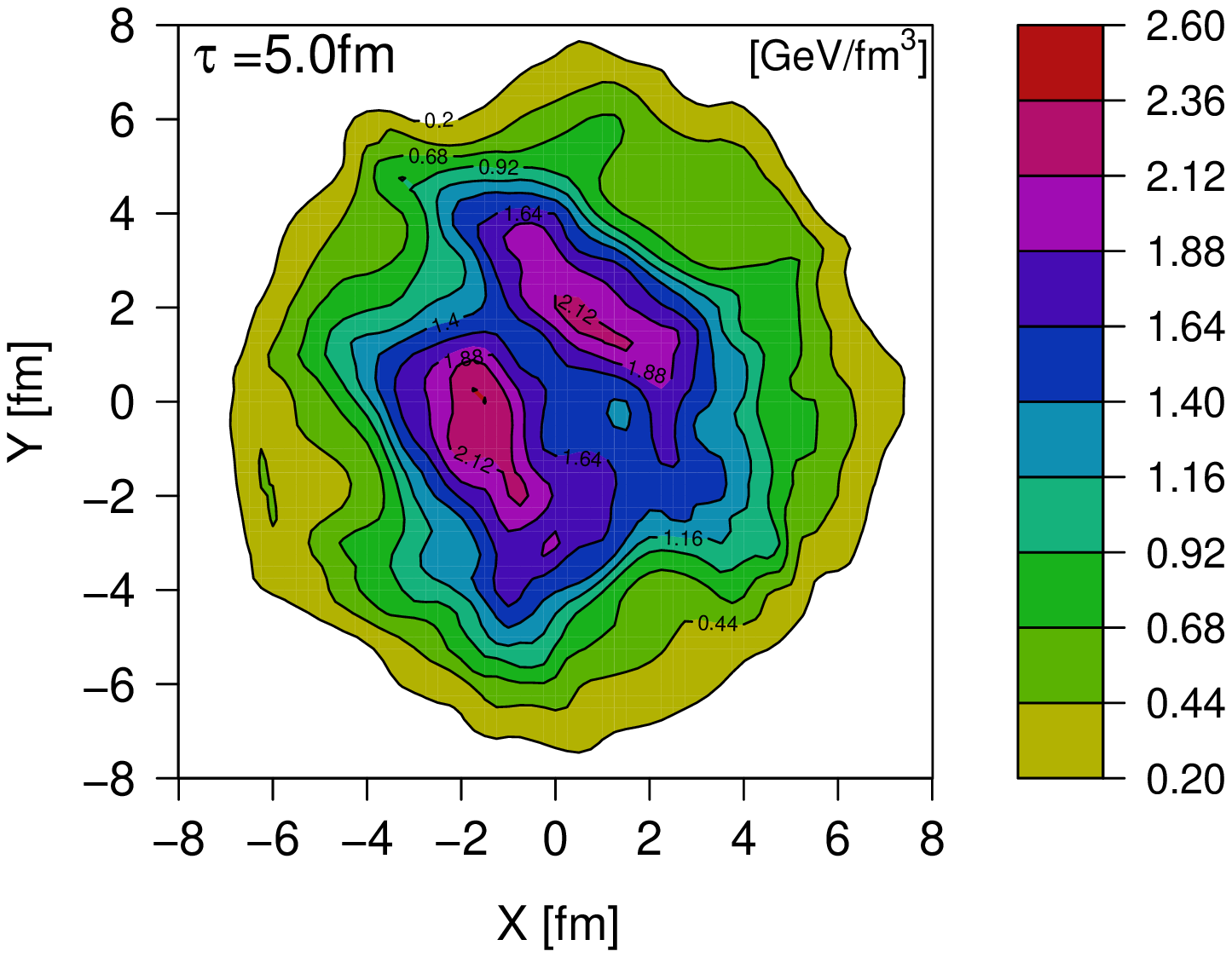}
\includegraphics[scale=0.33]{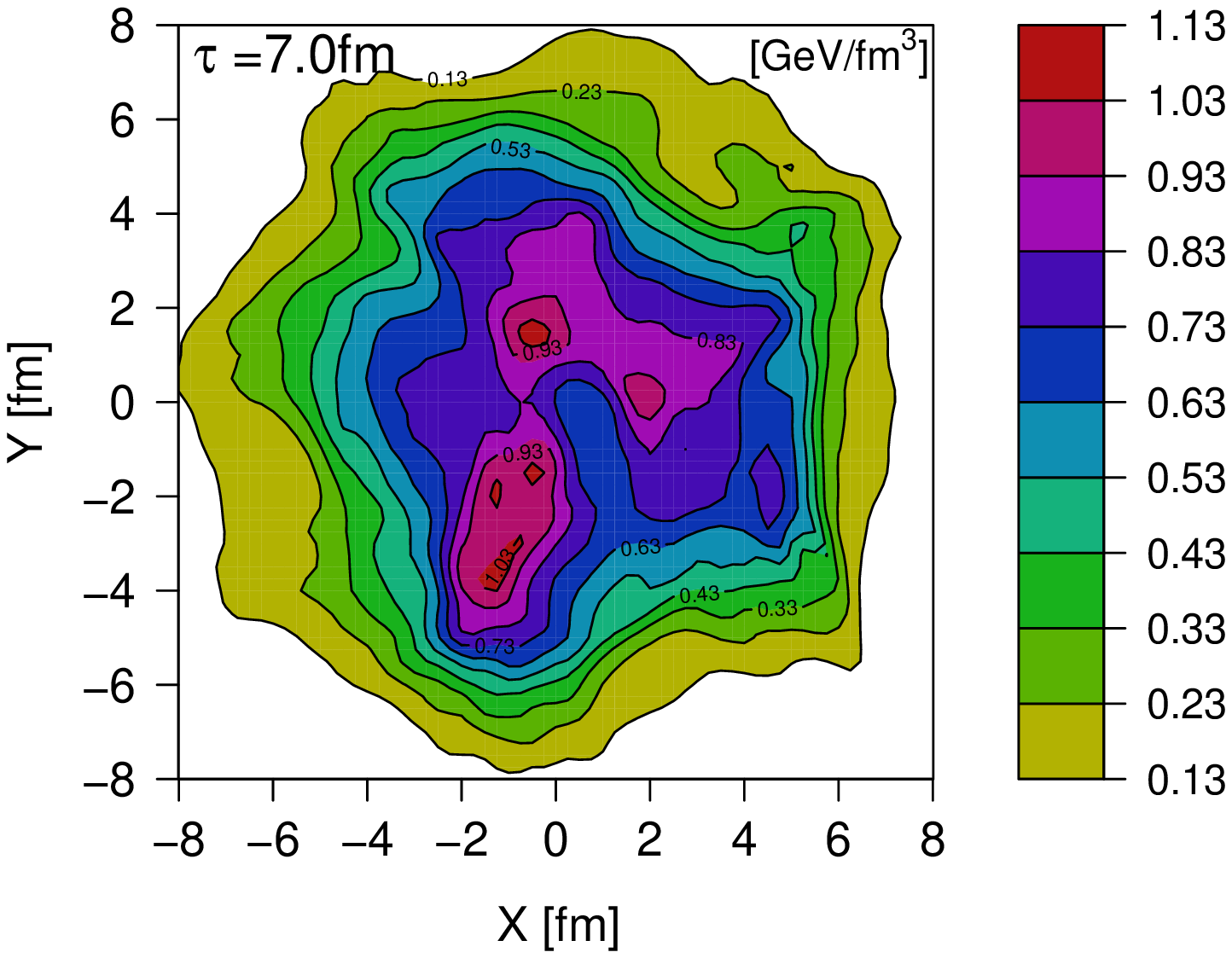}
\includegraphics[scale=0.33]{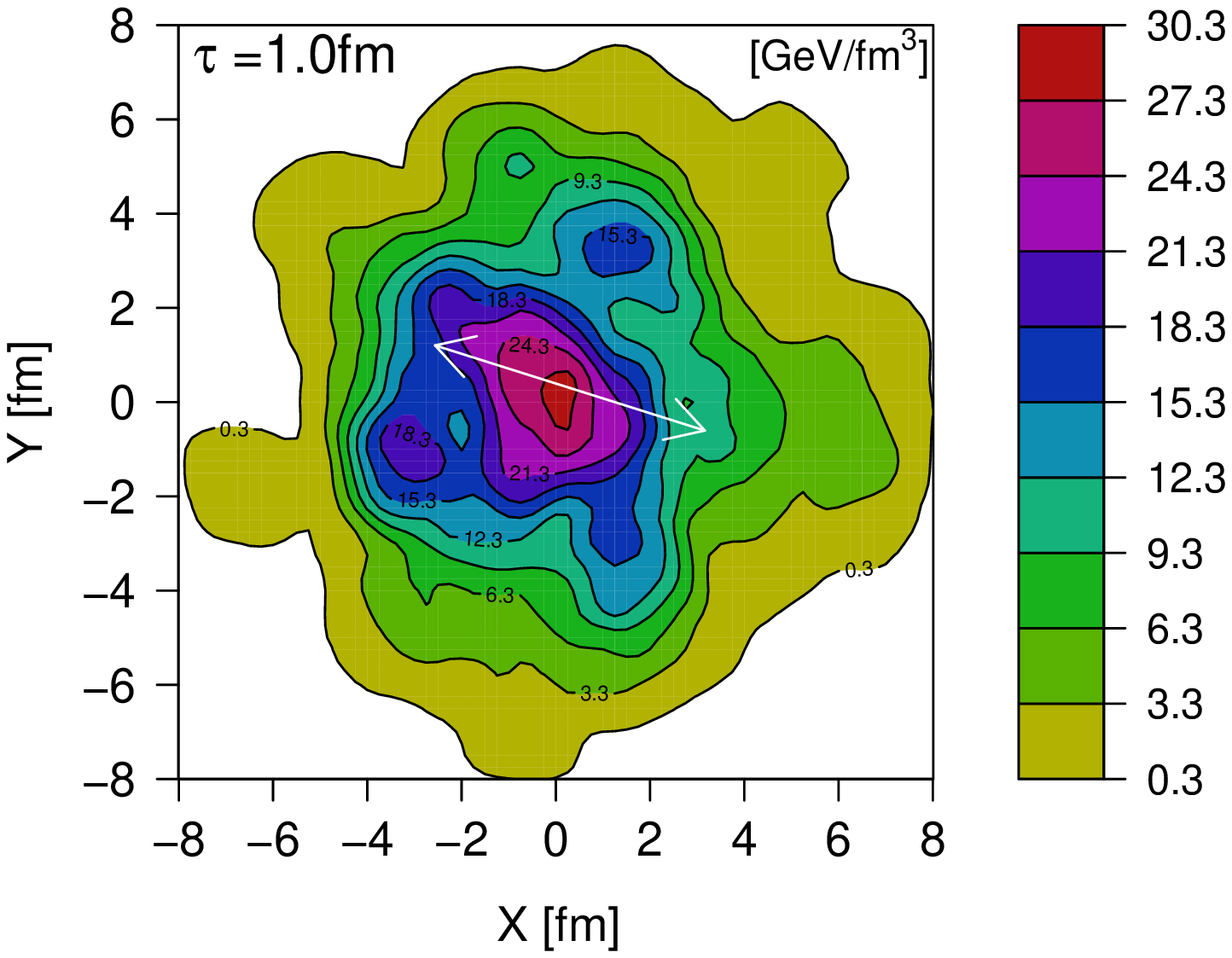}
\includegraphics[scale=0.33]{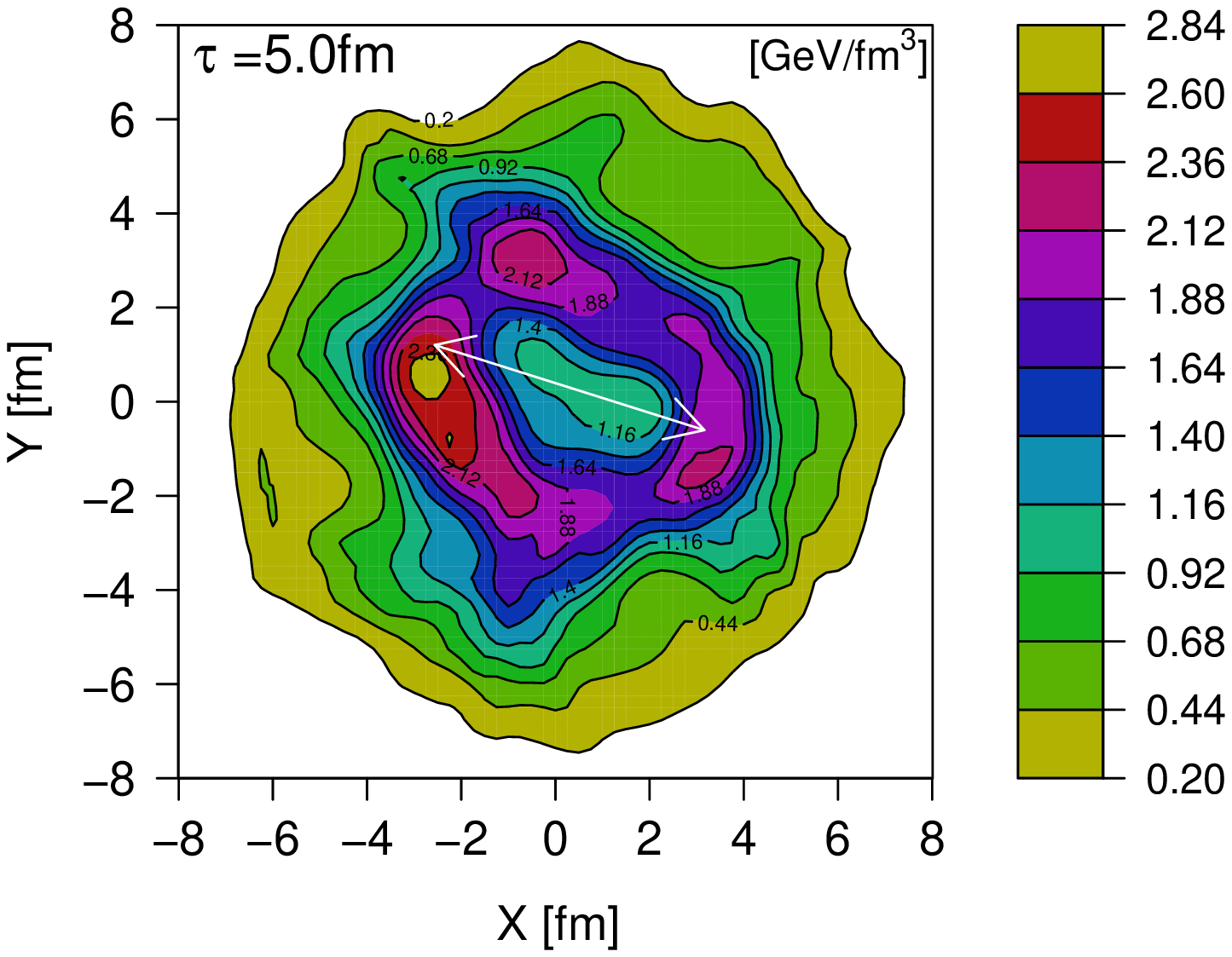}
\includegraphics[scale=0.33]{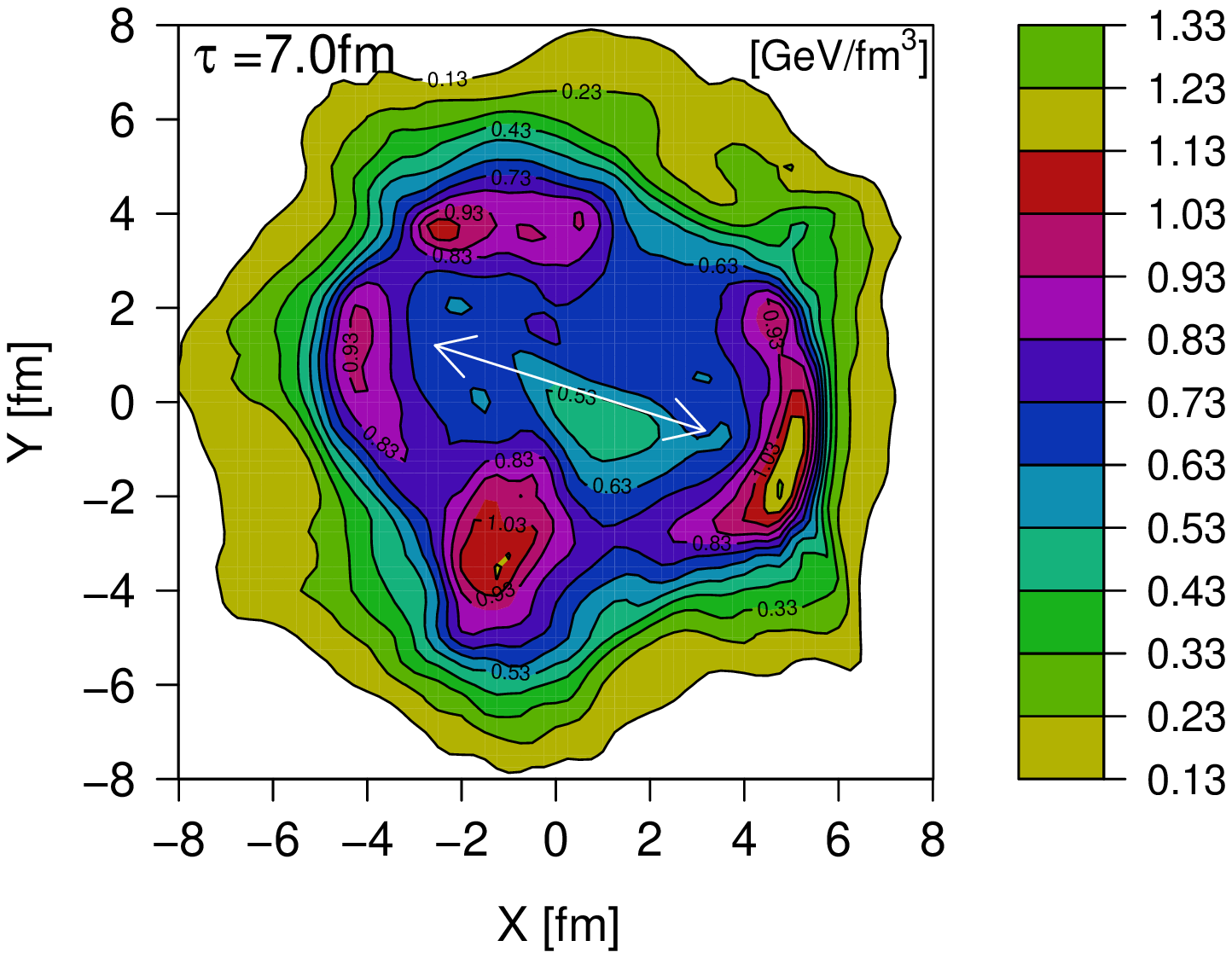}
\caption{\label{fig:ed_jet} (Color online) Hydrodynamic evolution, in the transverse plane at mid-rapidity, of a single event with (lower panels) and without (upper panels) the influence of the dijet. The initial energy density distribution corresponds to a randomly chosen central Au+Au collision at 200A GeV, computed using an implementation of the Monte Carlo Glauber model \cite{Drescher:2006ca,Drescher:2007ax}. The initial position of the dijet (lower panels) is indicated by the back-to-back arrows. In this computation, $\left. dE/d l \right|_0 = 20$ GeV/fm.}
\end{figure*}

In order to get an idea about the effect of the dijets in the medium, we show in Fig.\ \ref{fig:ed_jet} the hydrodynamic evolution, in the transverse plane at mid-rapidity, of a single event with (lower panels) and without (upper panels) the influence of the dijet. The initial energy density distribution corresponds to a randomly chosen Au+Au collision in the $(0-5)\%$ centrality window at 200A GeV, computed using an implementation of the Monte Carlo Glauber model \cite{Drescher:2006ca,Drescher:2007ax} and used throughout this paper. The distribution is quite irregular showing several regions where the energy is considerably concentrated (the so-called hot spots). The initial anisotropy in this model arises basically from the random fluctuating position of the incident nucleons; the regions where the energy density is large corresponds to the regions where the nucleon density is large. We normalize this distribution so that the maximum of the average temperature distribution coincides with the temperature of $0.31$ GeV (similar values can be found in the literature; see, for instance, Ref. \cite{Luzum:2010ae}).

In Fig.\ \ref{fig:ed_jet} (lower panels), the initial position of the dijet is indicated by the back-to-back arrows. Observe that the effect of the propagation of each jet in the medium is to pile up and accelerate the matter along its trajectory. As a consequence, the region behind the jet becomes less dense. In comparison with the event without dijets (upper panels), the hydrodynamic evolution is quite different. In this particular example, $\left. dE/d l \right|_0 = 20$ GeV/fm.

\begin{figure}[ht]
\includegraphics[scale=0.50]{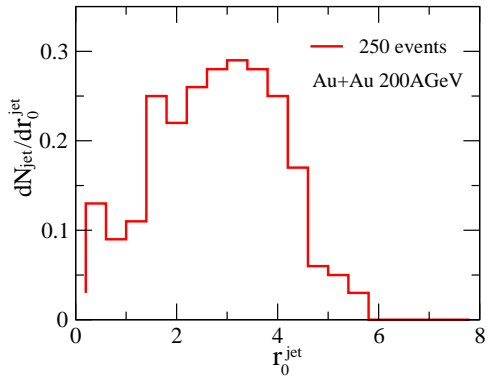}% Here is how to import EPS art
\caption{\label{fig:jet_e_2} (Color online) Distribution of the dijet initial distance, $r^{\hspace{0.5mm}\operatorname{jet}}_0$ (at $\tau=\tau_0$), with respect to the origin for 250 events (generated by the Monte Carlo Glauber model \cite{Drescher:2006ca,Drescher:2007ax}).}
\end{figure}

As suggested in Fig.\ \ref{fig:ed_jet} (lower panels), for each event we choose one hot spot as the initial position of the dijet (the dijet azimuthal angle is isotropic). This choice is weighted by the energy density at the hot spot position. Thus, the closer to the origin the hot spot is, the greater the probability of finding a dijet at its position becomes. In Fig.\ \ref{fig:jet_e_2}, we show the distribution of the dijet initial distance, $r^{\hspace{0.5mm}\operatorname{jet}}_0$ (at $\tau=\tau_0$), with respect to the origin for 250 events. As one can see, the distribution has a maximum at $r^{\hspace{0.5mm}\operatorname{jet}}_0 \sim 3$ fm. This happens because the probability of choosing one of the hot spots around the origin is greater than the probability of choosing only one hot spot at the origin. One expects that the dijets created with such a distribution on the transverse plane would also enhance the direct and triangular flows, besides the elliptic flow.

\begin{figure}[ht]
\includegraphics[scale=0.50]{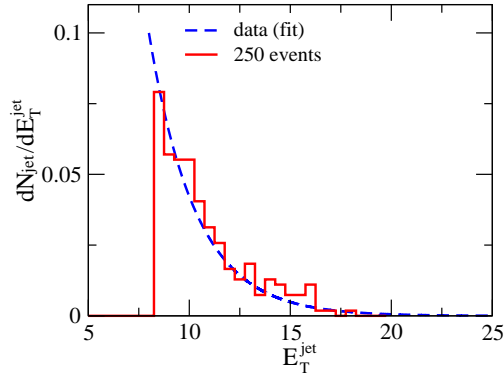}% Here is how to import EPS art
\caption{\label{fig:jet_e_1} (Color online) Jet yield per event as a function of the jet transverse energy, $E_{T}^{\hspace{0.5mm}\operatorname{jet}}$, for proton-proton collisions scaled by the number of binary collisions in Au+Au collisions at 200A GeV in the $(0-10)\%$  centrality window. The dashed line was obtained from RHIC data \cite{Salur:2008hs}. The solid line corresponds to the ensemble of 250 events used in this paper.}
\end{figure}

We assume that after losing their total energy the jets are absorbed by the medium. On the other hand, the
hadrons created by the fragmentation of the jets that have enough energy to escape the bulk nuclear matter are not included in the current version of the model. The implementation in our hydrodynamic code of fragmentation procedures is in progress. It is important to emphasize, as pointed out in Section \ref{sec:source}, that we are not taking into account the variation of the velocity of the  partonic jet during its motion through the medium. If the jet is fully absorbed, the velocity of the parton changes abruptly to the velocity of the fluid (see Ref. \cite{Betz:2008ka} for a study of stopped jets in hydrodynamics). In Fig.\ \ref{fig:jet_e_1}, we show the jet yield per event as a function of the jet transverse energy, $E_{T}^{\hspace{0.5mm}\operatorname{jet}}$, for proton-proton collisions scaled by the number of binary collisions in Au+Au collisions at 200A GeV in the $(0-10)\%$  centrality window. The dashed line was obtained from RHIC data \cite{Salur:2008hs}. The solid line corresponds to the ensemble of 250 events used in this paper. In our approach, we consider that $E_{T}^{\hspace{0.5mm}\operatorname{jet}}$ is the total energy of each parton of the dijet. Integrating this curve one finds that $N_{\hspace{0.5mm}\operatorname{jet}}\sim0.23$ (dijets per event). Because we are studying the $(0-5)\%$ centrality window, we have to correct this normalization using the factor $\left<N_{coll} \right>^{(0-5)\%}/\left<N_{coll} \right>^{(0-10)\%}$, which leads to $N_{\hspace{0.5mm}\operatorname{jet}}\sim0.25$, where $\left<N_{coll}\right>$ is the average number of binary collisions.

Summarizing, the procedure to compute an observable event-by-event, including the jet parametrization, is the following: (i) the initial conditions are computed using an implementation of the Monte Carlo Glauber model \cite{Drescher:2006ca,Drescher:2007ax}; (ii) one hot spot of the event is chosen as the initial position of the dijet (one dijet per event), taking as weights the energy density at the hot spot positions (the azimuthal angle of the dijet is isotropic); (iii) the total energy of each jet (the same for both jets in the pair) is chosen according to the jet yield per event (see Fig.\ \ref{fig:jet_e_1}); (iv) the hydrodynamic evolution is computed through the SPH method \cite{Aguiar:2000hw,Andrade:2013poa} and (v) the final spectra (for direct positively charged pions) is computed using the Cooper-Frye prescription \cite{Cooper:1974mv}. At the end of the simulation, the average value of a given observable is calculated over an ensemble of events. We call ``mixed ensemble" the ensemble composed by 1000 events, following the proportion fixed by the jet yield per event, i.e., 750 events computed without dijets and 250 with dijets. On the other hand, the ``jet ensemble" includes only events with dijets (250 events).

\section{\label{sec:results} Results}

\begin{figure}[ht]
\includegraphics[scale=0.42]{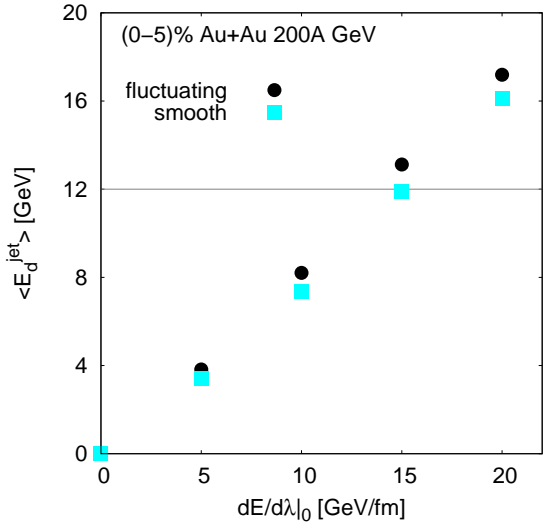}% Here is how to import EPS art
\caption{\label{fig:jet_ael} (Color online) Average energy deposited in the medium, around mid-rapidity, by the dijet, $< E_{d}^{\hspace{0.5mm}\operatorname{jet}}>$, in the $(0-5)\%$ centrality window, as a function of the reference energy loss rate $\left. dE/d l \right|_0$ (using the jet ensemble).}
\end{figure}

In Fig.\ \ref{fig:jet_ael} we show the average energy deposited in the medium, around the mid-rapidity, by the dijet, $< E_{d}^{\hspace{0.5mm}\operatorname{jet}}>$, in the $(0-5)\%$ centrality window, as a function of the reference energy loss rate $\left. dE/d l \right|_0$ (using the jet ensemble). To compute the curve labeled ``smooth" (squares) we replace, in each event, the fluctuating initial energy density distribution by a smooth one while keeping unchanged the initial position of the dijet. Also, one can see that the fluctuations slightly enhance the suppression of jets in the medium. Note that the magnitude of $< E_{d}^{\hspace{0.5mm}\operatorname{jet}}>$, using our upper limit for the coupling between the QGP and the jets, $\left. dE/d l \right|_0=20$ GeV/fm, is relatively small, on the order of $16$ GeV (8 GeV for each jet of the pair, on average). This is mainly because of the violent longitudinal expansion that quickly rarefies the QGP and therefore quickly enhances the mean free path of the partonic jets. For $\left. dE/d l \right|_0=5$GeV/fm, one finds that $< E_{d}^{\hspace{0.5mm}\operatorname{jet}}>$ is slightly smaller than 4 GeV.

\begin{figure}[ht]
\includegraphics[scale=0.42]{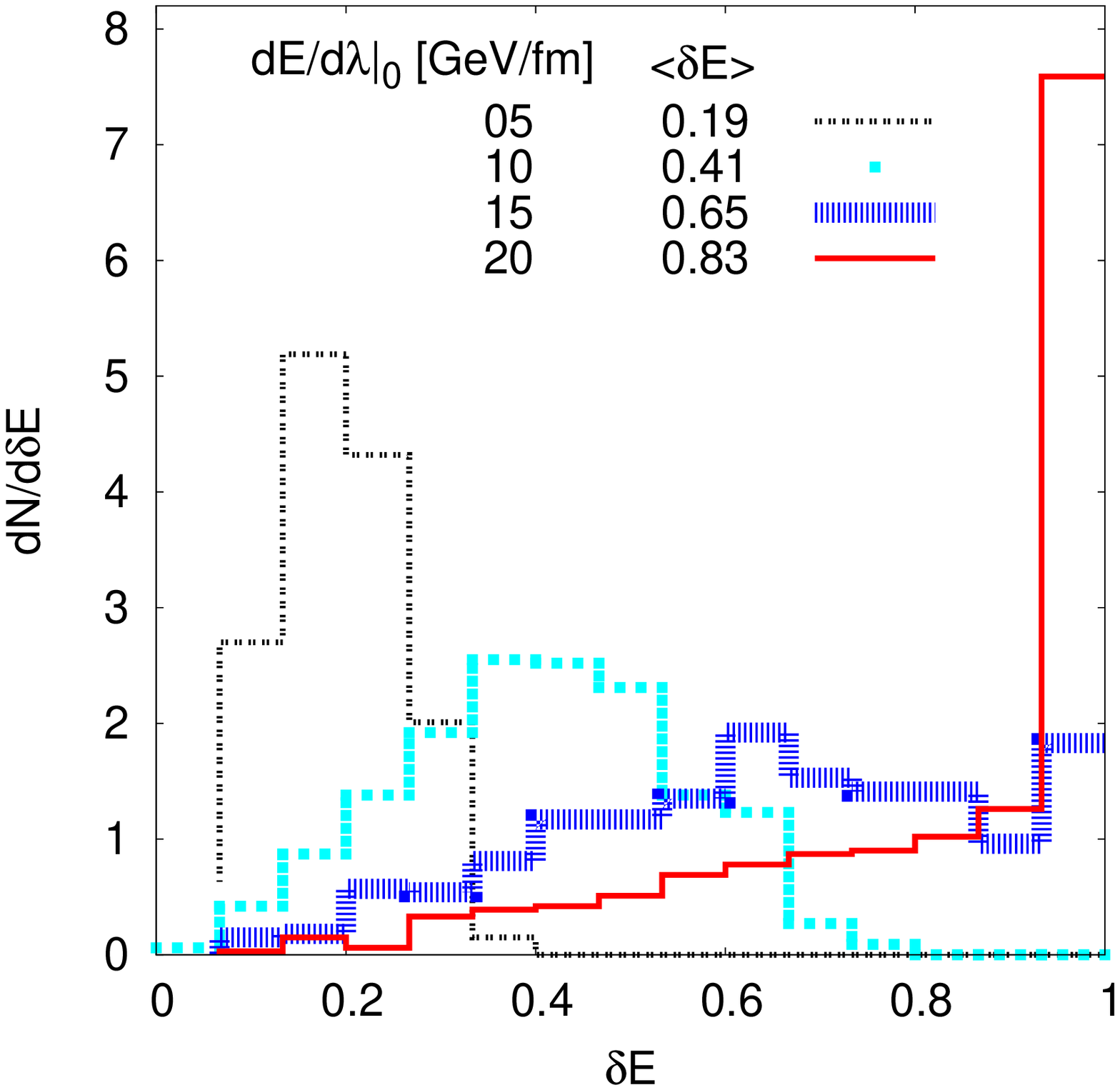}% Here is how to import EPS art
\caption{\label{fig:jhisto_ef2} (Color online) Distribution of the ratio $\delta E = E_{d}^{\hspace{0.5mm}\operatorname{jet}}/ E_{T}^{\hspace{0.5mm}\operatorname{jet}}$ for four values of the parameter $\left. dE/d l \right|_0$. The respective average value $<\delta E>$ is shown on the plot.}
\end{figure}

In Fig.\ \ref{fig:jhisto_ef2}, we show the distribution of the ratio $\delta E = E_{d}^{\hspace{0.5mm}\operatorname{jet}}/ E_{T}^{\hspace{0.5mm}\operatorname{jet}}$, i.e., the relative amount of energy (with respect to the initial jet transverse energy) that is lost to the medium. This distribution is shown for four values of the parameter $\left. dE/d l \right|_0$. The respective average value $<\delta E>$ is shown on the plot. Observe that $<\delta E>$ gets close to unity when $\left. dE/d l \right|_0$ is increased. In fact, depending on the magnitude of the coupling between the jets and the QGP,  a considerable fraction of the jets may be completely absorbed by the medium. These distributions survey, in our model, an estimative of the suppression of the jets in the medium and can be used to calibrate the free parameter $\left. dE/d l \right|_0$. As we are going to see in the next plot, for $\left. dE/d l \right|_0 \gtrsim 15$ GeV/fm, which corresponds to a suppression on average greater than $65\%$,  the jet quenching effect may create relevant additional anisotropic flow.

\begin{figure*}[ht]
\includegraphics[scale=0.45]{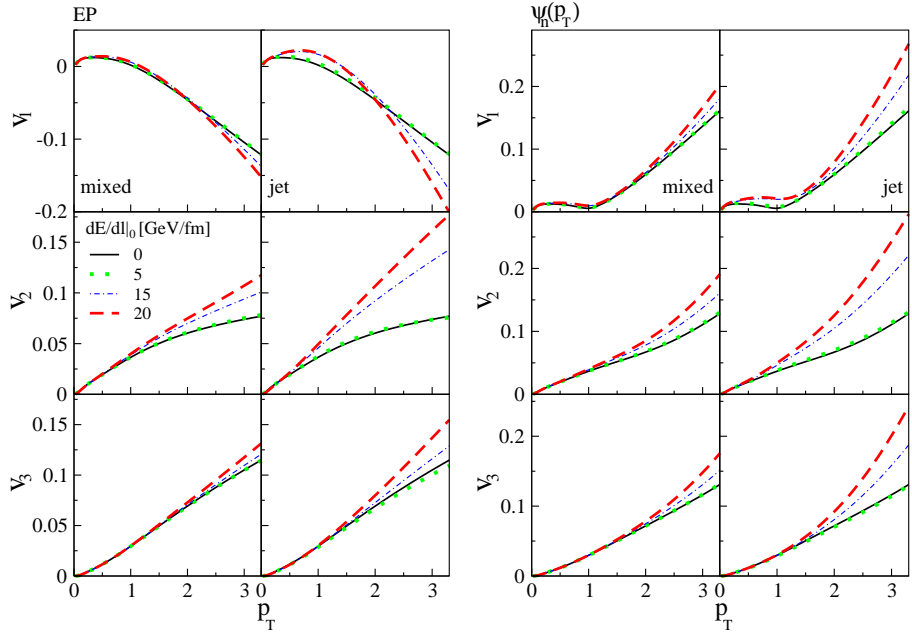}% Here is how to import EPS art
\caption{\label{fig:vn_3} (Color online) Transverse momentum dependence of the $v_n$ coefficients ($n=1,2,3$) for four values of the parameter $\left. dE/d l \right|_0$. The left panels correspond to the event plane method. On the right panels we show the same observables computed using $\Psi_n=\Psi_n \left( p_T\right)$, i.e., the phase is computed for each $p_T$ bin \cite{Andrade:2013poa}. The panels labeled ``mixed" correspond to an ensemble of 1000 events that includes 750 events without and 250 events with dijets. The panels labeled ``jet" correspond to an ensemble of 250 events that includes only events with dijets.}
\end{figure*}

In Fig.\ \ref{fig:vn_3}, we show our results for the transverse momentum dependence of the $v_n$ coefficients ($n=1,2,3$) for four values of the parameter $\left. dE/d l \right|_0$. The left panels correspond to the Event Plane method (EP) where the phase $\Psi_n$ is computed using all the hadrons of the event \cite{Poskanzer:1998yz}. In the right panels, we show the same observables computed using $\Psi_n=\Psi_n \left( p_T\right)$, i.e., the phase is computed for each $p_T$ bin (see the necessary details in \cite{Andrade:2013poa}). Consequently, the latter procedure maximizes the anisotropy. The negative sign observed in the coefficient $v_1\left(p_T \right)$, computed using the event plane method, is a consequence of momentum conservation: if the low $p_T$ particles move in one direction, the higher $p_T$ particles must move in the opposite direction to conserve momentum. As one can see, in the majority of the cases, the effects of the jets are important in the region of intermediate $p_T$ ($p_T\gtrsim1$GeV). However, using our lower limit for the coupling between the QGP and the jets, $\left. dE/d l \right|_0=5$GeV/fm, the results are nearly identical to the results without jets. Finally, the anisotropy is enhanced, as expected, when one includes only events with jets (the jet ensemble).

In this paper, we use the following definition for the $n$th eccentricity (see, for instance, Ref. \cite{Teaney:2010vd})

\begin{equation}
\varepsilon_{m,n} = \frac{\left\{r^m \cos \left[n \left(\phi - \Phi_{m,n} \right)  \right]  \right\}}{\left\{r^m \right\}}\,,
\label{eq:eccentricity}
\end{equation}

\noindent
where

\begin{equation}
\Phi_{m,n}=\frac{1}{n} \tan^{-1} \left( \frac{\left\{r^m \sin \left(n \phi \right) \right\}}{\left\{r^m \cos \left(n \phi \right) \right\}} \right),
\label{eq:eccentricity_phase}
\end{equation}

\noindent
$r^m = \left(x^2+y^2\right)^\frac{m}{2}$ and $\phi=\tan^{-1} \left(y/x \right)$. The notation $\{...\}$ indicates the average weighted by the energy density profile in the transverse plane (see Fig.\ \ref{fig:ed_jet}, for $\tau=1$ fm, upper panels). For the sake of simplicity and following the original proposal discussed in Ref. \cite{Alver:2010gr}, we set $m=2$. Particularly, we choose the solution of Eq. (\ref{eq:eccentricity_phase}) that makes the eccentricity $\varepsilon_{2,n}$ a positive quantity. For instance, the phase $\Phi_{2,2}$ corresponds to the major axis of the ellipse.

\begin{figure*}
\includegraphics[scale=0.7]{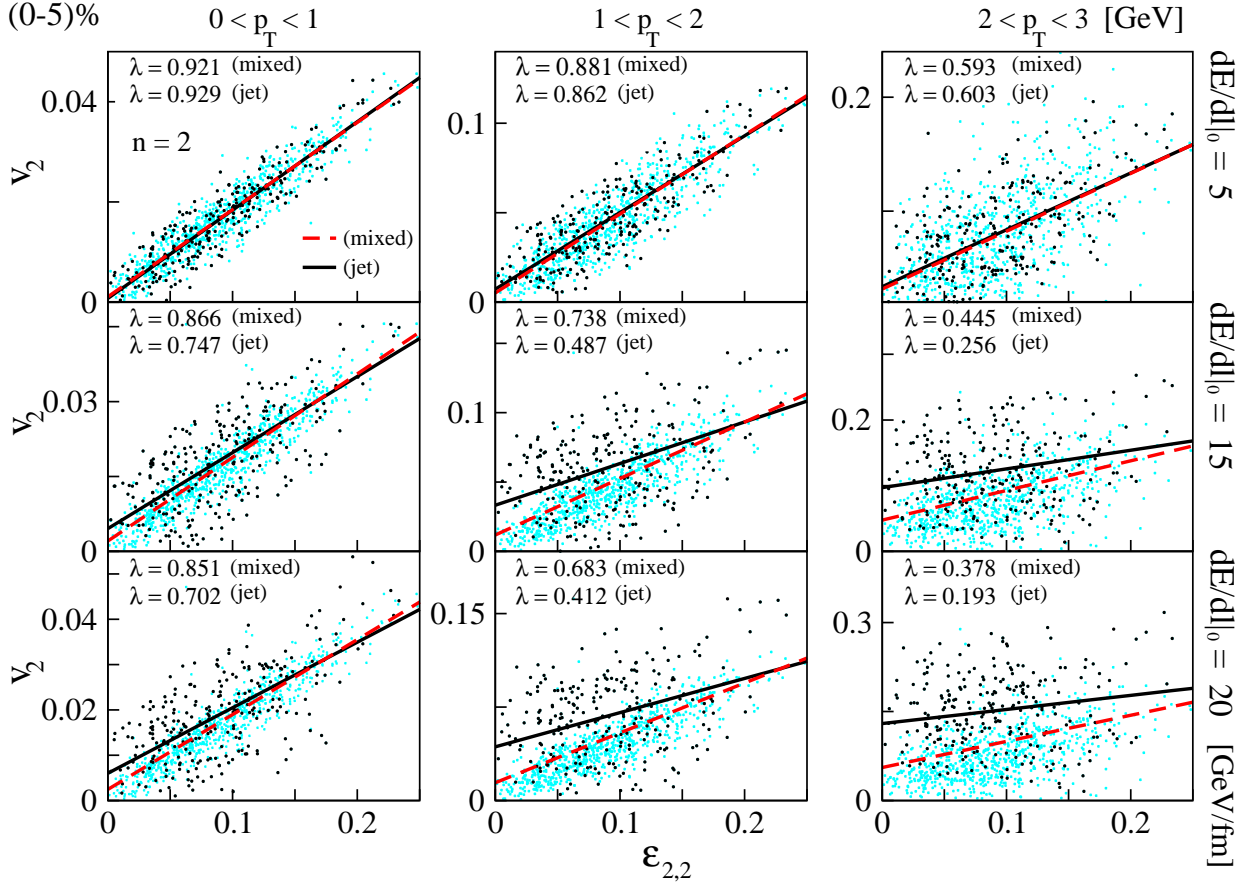}% Here is how to import EPS art
\caption{\label{fig:v2_x_e22_0_5} (Color online) Correlation between the eccentricity $\varepsilon_{2,2}$ and the flow coefficient $v_2$, for three values of the parameter $\left. dE/d l \right|_0$. The dashed lines correspond to linear fits computed using the mixed ensemble (black and light dots). The solid lines were computed using the jet ensemble (black dots). Three ranges of transverse momentum are presented. $\lambda$ is the linear correlation coefficient.}
\end{figure*}

\begin{figure*}
\includegraphics[scale=0.7]{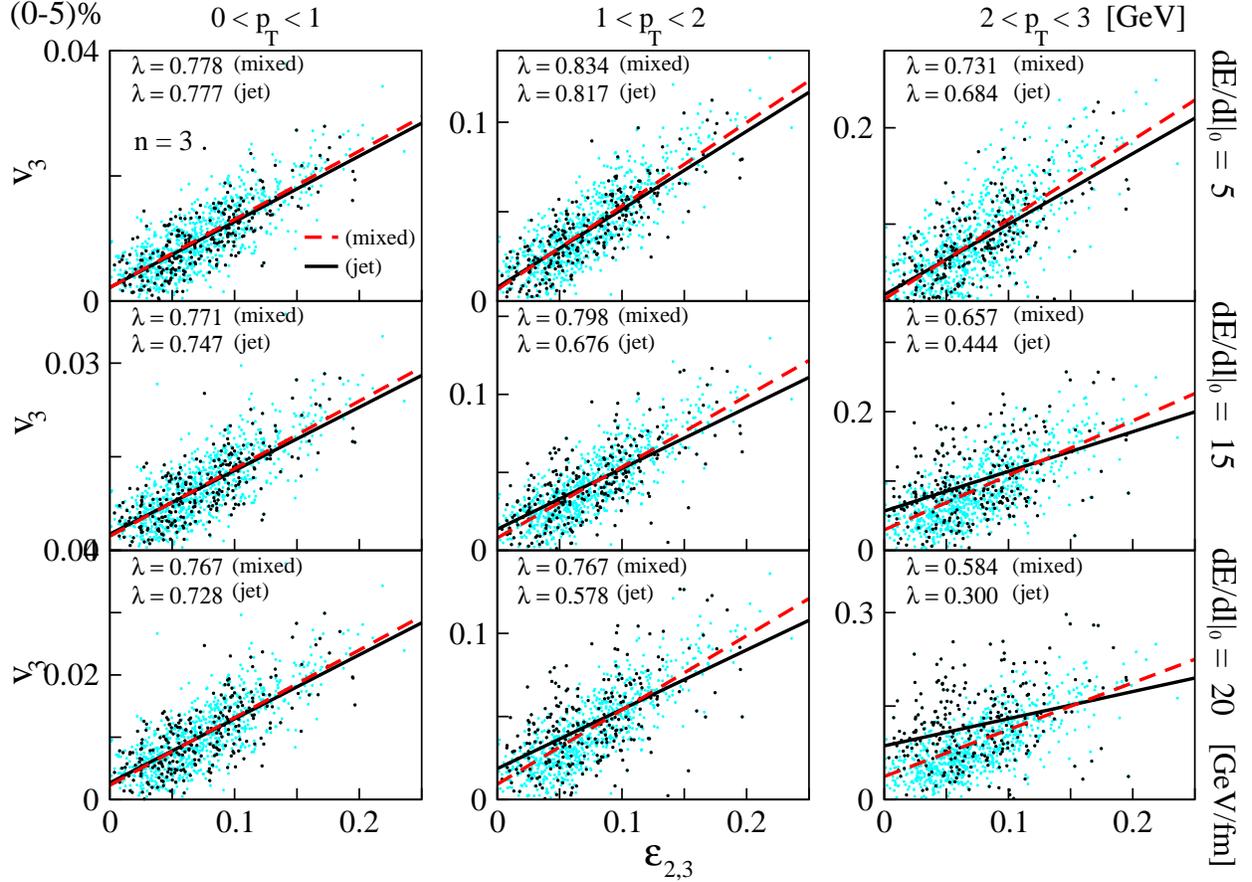}% Here is how to import EPS art
\caption{\label{fig:v3_x_e23_0_5} (Color online) Correlation between the eccentricity $\varepsilon_{2,3}$ and the flow coefficient $v_3$, for three values of the parameter $\left. dE/d l \right|_0$. The dashed lines correspond to linear fits computed using the mixed ensemble (black and light dots). The solid lines were computed using the jet ensemble (black dots). Three ranges of transverse momentum are presented. $\lambda$ is the linear correlation coefficient.}
\end{figure*}

In Fig.\ \ref{fig:v2_x_e22_0_5}, we show the correlation between the eccentricity $\varepsilon_{2,2}$ and the flow coefficient $v_2$, for three values of the parameter $\left. dE/d l \right|_0$. The dashed lines correspond to linear fits computed using the mixed ensemble (black and light dots). The solid lines were computed using the jet ensemble (black dots). Three ranges of transverse momentum are presented. A similar graph is shown in Fig.\ \ref{fig:v3_x_e23_0_5}, for the correlation between $\varepsilon_{2,3}$ and $v_3$. Moreover,

\begin{equation}
\lambda_n=\frac{\left< \left(\varepsilon_{2,n} -  \left< \varepsilon_{2,n} \right> \right) \left(v_n -  \left< v_n \right> \right)  \right> }{ \sqrt{\left< \left(\varepsilon_{2,n} -  \left< \varepsilon_{2,n} \right> \right)^2 \right> \left< \left(v_n -  \left< v_n \right> \right)^2 \right> }}
\label{equ:lcoeff}
\end{equation}

\noindent
is the linear correlation coefficient ($n=2,3$). The closer to the unit $\left| \lambda_n \right|$ is, the stronger the linear correlation between the variables $\varepsilon_{2,n}$ and $v_n$ becomes. In fact, when $\lambda\sim 1$  ($\lambda\sim -1$) both variables show a strong linear correlation (anti-correlation).

One can see that in the low $p_T$ region, $0<p_T<1$GeV, the influence of jets in the hydrodynamic evolution of the QGP is negligible since the results obtained in this range are similar to the ones computed using events without jets (see Ref. \cite{Andrade:2013poa}). On the other hand, for $2<p_T<3$GeV, the presence of jets reduce considerably the correlation between the Fourier coefficient $v_n$ and the eccentricity $\varepsilon_{2,n}$ ($n=2,3$). For instance, using $\left. dE/d l \right|_0=20$GeV/fm, one observes that $\lambda_2=0.378$ for the mixed ensemble. Note that the anisotropic flow created by the jets can be clearly seen in events with zero eccentricity (the linear fit crosses the vertical axis above the origin).

\begin{figure*}
\includegraphics[scale=0.55]{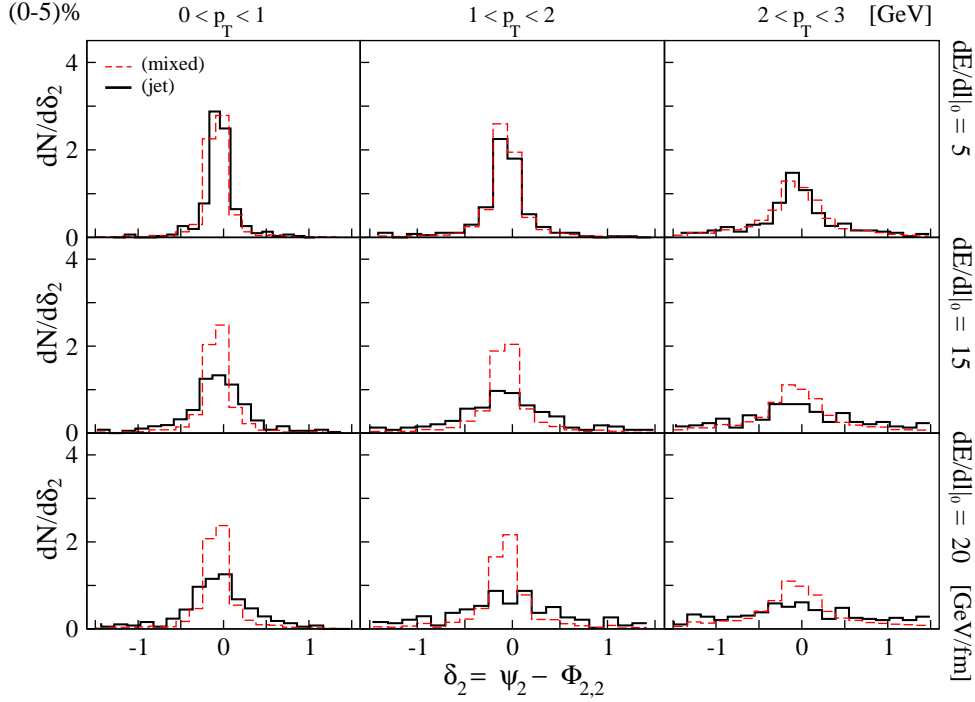}% Here is how to import EPS art
\caption{\label{fig:dN_dd2_x_d2_0_5_jet} (Color online) Distribution of the phase difference $\delta_2 = \Psi_2 - \Phi_{2,2}$, for three values of the parameter $\left. dE/d l \right|_0$. The dashed lines correspond to the mixed ensemble and the solid lines to the jet ensemble. Three ranges of transverse momentum are presented. Note that $\Psi_2$ is rotated by $\pi/2$ in order to achieve the smallest angular difference with respect to the angle $\Phi_{2,2}$. All distributions are normalized to unity.}
\end{figure*}

\begin{figure*}
\includegraphics[scale=0.55]{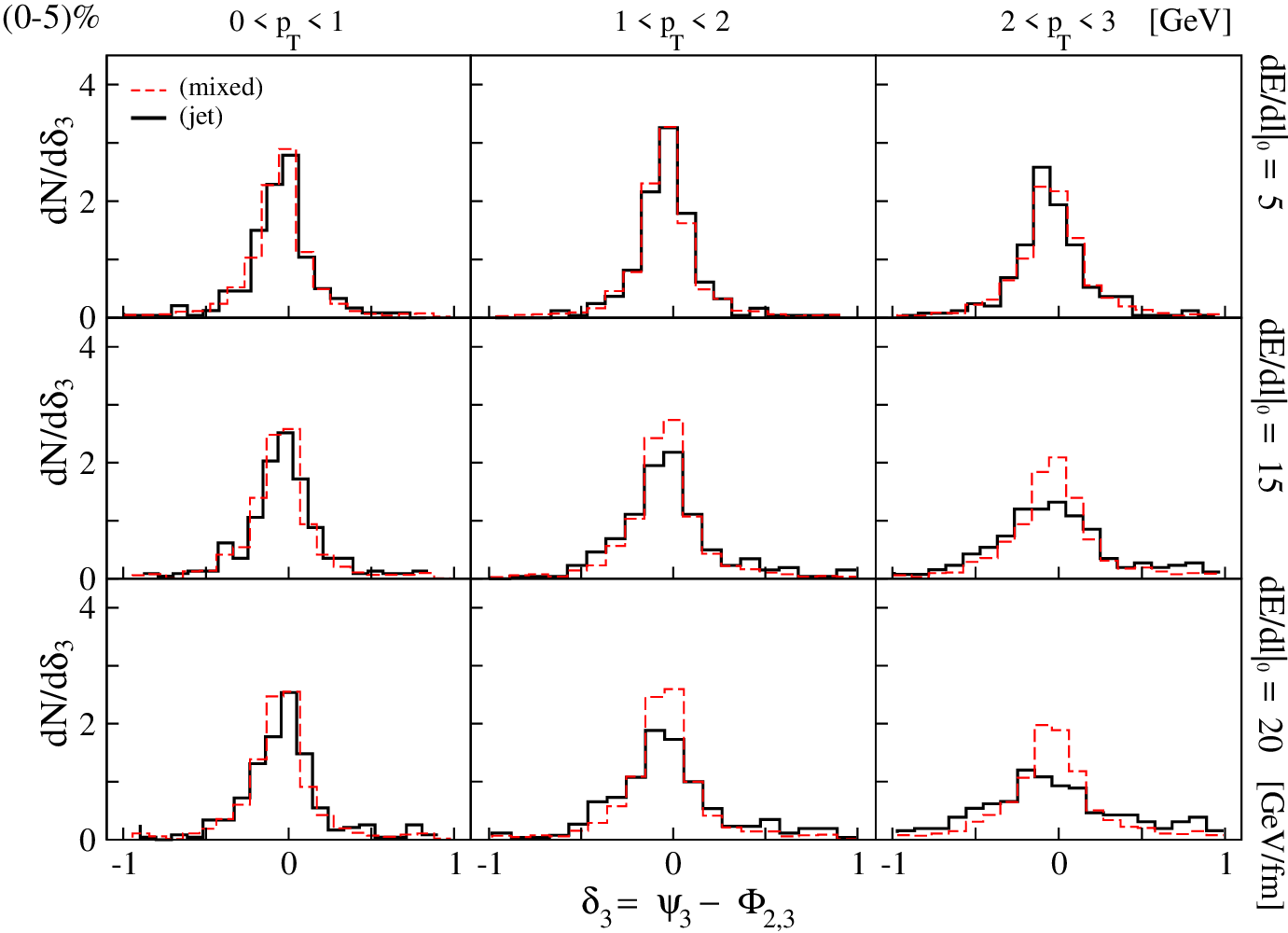}% Here is how to import EPS art
\caption{\label{fig:dN_dd3_x_d3_0_5_jet} (Color online) Distribution of the phase difference $\delta_3 = \Psi_3 - \Phi_{2,3}$, for three values of the parameter $\left. dE/d l \right|_0$. The dashed lines correspond to the mixed ensemble and the solid lines to the jet ensemble. Three ranges of transverse momentum are presented. Note that $\Psi_3$ is rotated by $\pi/3$ in order to achieve the smallest angular difference with respect to the angle $\Phi_{2,3}$. All distributions are normalized to unity.}
\end{figure*}

In Fig.\ \ref{fig:dN_dd2_x_d2_0_5_jet}, we show the distribution of the phase difference $\delta_2 = \Psi_2 - \Phi_{2,2}$, for three values of the parameter $\left. dE/d l \right|_0$. The dashed lines correspond to the mixed ensemble and the solid lines to the jet ensemble. Three ranges of transverse momentum are presented. A similar graph is shown in Fig.\ \ref{fig:dN_dd3_x_d3_0_5_jet} for the distribution of the phase difference $\delta_3 = \Psi_3 - \Phi_{2,3}$. Note that here $\Psi_n$ is rotated by $\pi/n$ in order to achieve the smallest angular difference with respect to the angle $\Phi_{2,n}$. All distributions are normalized to the unit. Clearly, the effect of jets is to make the distributions wider, mainly in the region where $2<p_T<3$GeV. However, there is still an excess of events close to the origin. Comparing the distributions computed using the mixed and jet ensembles, the latter are broader, as expected. In the lower $p_T$ region, for the mixed ensemble, the distributions are similar to the ones obtained from events without jets \cite{Andrade:2013poa}, i.e., they show a sharp peak at the origin.

Following Ref. \cite{Andrade:2013poa}, the azimuthal component of the dihadron correlation function, $C\left(\Delta \phi \right)$, can be expanded in terms of the pair  $\left\{v_n,\Psi_n \right\}$. Thus, one finds that

\begin{equation}
C\left(\Delta \phi \right)=  c_0 + \sum_n c_n \cos(n \Delta \phi) + \sum_n \tilde{c}_n \sin(n \Delta \phi)
\label{eq:2_part_cf_fo}
\end{equation}

\noindent
where

\begin{equation}
c_0 =  \frac{\left<v_{0}^a v_{0}^t\right>}{\left<v_{0}^t\right>},
\label{eq:c0}
\end{equation}

\begin{equation}
c_n = \frac{2}{\left<v_{0}^t\right>}   \left<v_{0}^a v_{0}^t v_{n}^a v_{n}^t   \cos\left[n\left(\Psi_{n}^t - \Psi_{n}^a \right) \right] \right>,
\label{eq:cn}
\end{equation}

\noindent
and

\begin{equation}
\tilde{c}_n = \frac{2}{\left<v_{0}^t\right>} \left<v_{0}^a v_{0}^t v_{n}^a v_{n}^t  \sin\left[n\left(\Psi_{n}^a - \Psi_{n}^t \right) \right]\right>.
\label{eq:cnt}
\end{equation}

\noindent
The index $a$ and $t$ correspond to the associated particles and triggers, respectively. The brackets indicate the average over events (an arithmetic mean). In addition, it was shown in Ref. \cite{Andrade:2013poa} that the relative phase $\Delta_n = \Psi_{n}^t - \Psi_{n}^a$ is independent on the $v_n$ coefficients. Then, one expects for a sufficiently large number of events that

\begin{equation}
c_n \sim \frac{2}{\left<v_{0}^t\right>}   \left<v_{0}^a v_{0}^t v_{n}^a v_{n}^t \right>  \left< \cos\left[n\left(\Psi_{n}^t - \Psi_{n}^a \right) \right] \right>
\label{eq:cnf}
\end{equation}

\noindent
and similarly for the $\tilde{c}_n$ coefficients.

\begin{figure*}
\includegraphics[scale=0.55]{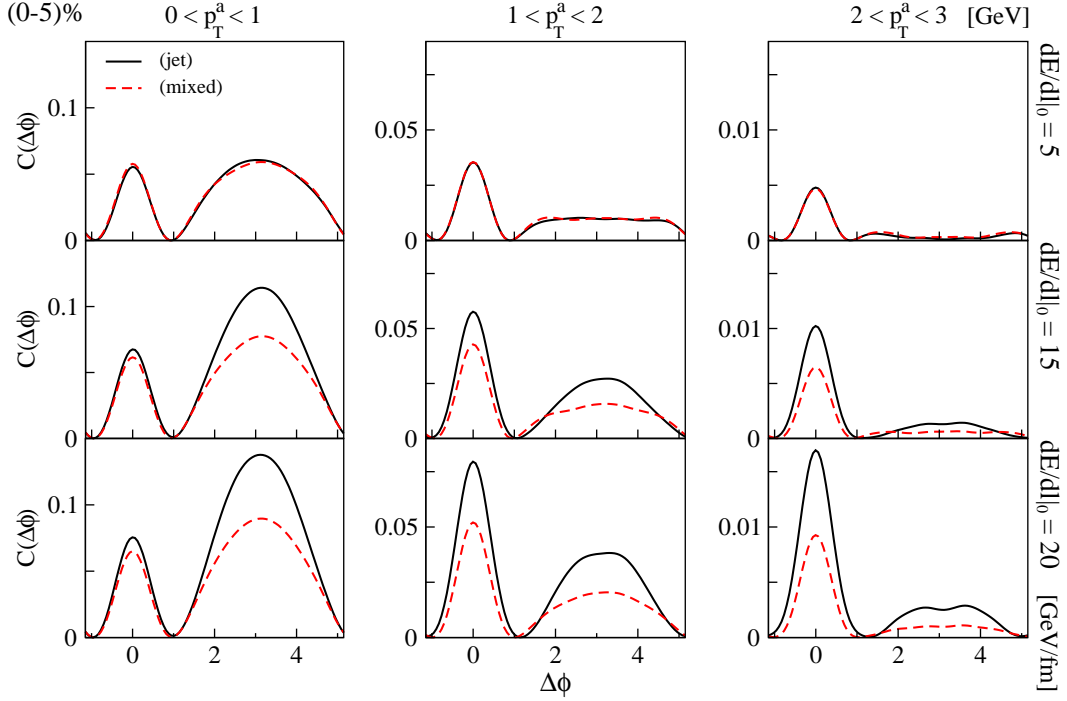}% Here is how to import EPS art
\caption{\label{fig:dN_ddphi_x_dphi_0_5_jet} (Color online) Azimuthal component of the dihadron correlation function $C \left(\Delta \phi \right)$, for three values of the parameter $\left. dE/d l \right|_0$. The dashed lines correspond to the mixed ensemble and the solid lines to the jet ensemble. Three ranges of transverse momentum of the associated particles are presented. The range in transverse momentum for the triggers is defined as $3 < p_T^{\mathbf{trigg}}< 5$ GeV.}
\end{figure*}

\begin{figure*}
\includegraphics[scale=0.55]{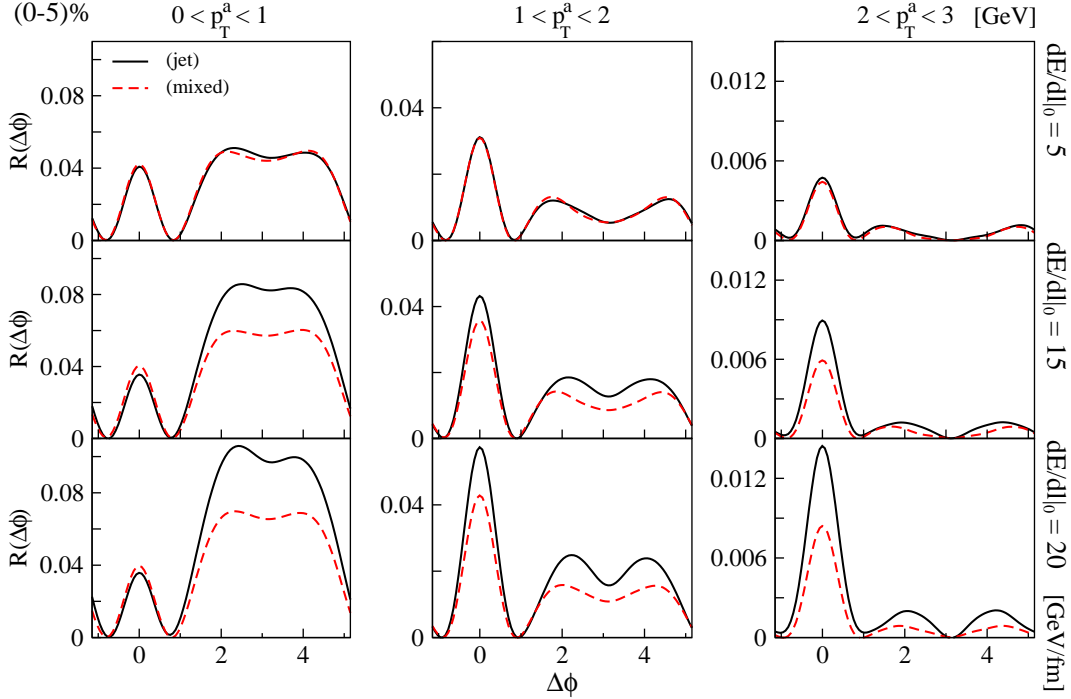}% Here is how to import EPS art
\caption{\label{fig:dN_ddphi_x_dphi_0_5_jet_2} (Color online) Azimuthal component of the background subtracted dihadron correlation function $R \left(\Delta \phi \right)$, for three values of the parameter $\left. dE/d l \right|_0$. The dashed lines correspond to the mixed ensemble and the solid lines to the jet ensemble. Three ranges of transverse momentum of the associated particles are presented. The range in transverse momentum for the triggers is defined as $3 < p_T^{\mathbf{trigg}}< 5$ GeV.}
\end{figure*}

In Fig.\ \ref{fig:dN_ddphi_x_dphi_0_5_jet}, we show the azimuthal component of the dihadron correlation function $C \left(\Delta \phi \right)$, for three values of the parameter $\left. dE/d l \right|_0$. The dashed lines correspond to the mixed ensemble and the solid lines to the jet ensemble. Three ranges of transverse momentum of the associated particles are presented. The range in transverse momentum for the triggers is defined as $3 < p_T^{\mathbf{trigg}}< 5$ GeV. The corresponding background subtracted function $R \left(\Delta \phi \right)$ is shown in Fig.\ \ref{fig:dN_ddphi_x_dphi_0_5_jet_2} (there is a small asymmetry due to the finite number of events in the jet ensemble that is more visible when the associated particles are in the first $p_T$ bin and the energy loss is maximal. This asymmetry will, of course, vanish if more events are considered. We note, however, the presence of such asymmetry does not change any of the conclusions drawn in this paper). 

The method used to remove the background and define the function $R \left(\Delta \phi \right)$ is a variation of the well known mixed event method (see \cite{Takahashi:2009na,Andrade:2013poa}). In this approach, the associated particles and the triggers are chosen in different events, producing a mixed correlation. This is commonly employed to remove the longitudinal correlation that comes from the shape of the longitudinal distribution of particles. Here, the events which will be mixed are aligned according to the direction of the event plane $\Psi_2^{\mathbf{EP}}$. Clearly, this procedure generates a background of the type $c_2^{\mathbf{mix}} \cos \left(2 \Delta \phi \right)$.

The effect of the jets on the profile of the dihadron angular correlation function is essentially to modify the relative height between the near-side peak and away-side peaks. This is a consequence of the direct flow, $v_1$, created by the dijet. Observe that the away-side peaks are higher in the region of low $p_T^a$ ($p_T^a<1$ GeV) and the situation is inverted in the region of intermediate $p_T^a$ $\left(2<p_T^a<3\right)$ GeV. The element that controls this behavior is the sign of the cosine of the average phase difference $\Delta_1=\Psi_{1}^t - \Psi_{1}^a$ [see formula (\ref{eq:cnf})]. In the region of low $p_T^a$,  $\left< \cos\Delta_1  \right> < 0$ and consequently $c_1 < 0$ (the away-side peaks are enhanced). On the other hand, in the region of intermediate $p_T^a$, $\left< \cos\Delta_1  \right> > 0$ and therefore $c_1 > 0$ (the near side peak is enhanced). The reason why the coefficient $c_1$ is not a positive definite function can be understood from the analysis of the curve $v_1 \left(p_T \right)$, computed through the event plane method (see Fig.\ \ref{fig:vn_3}, EP, mixed). One can see that the direct flow is positive for low $p_T$ and negative for intermediate and high $p_T$. Therefore, choosing the associated particles in the low $p_T$ region one finds that $\Delta_1 \sim \pi$. On the other hand, choosing the associated particles in the intermediate $p_T$ region one finds that $\Delta_1 \sim 0$. Observe that in the jet ensemble the effect of $v_1$ on the dihadron angular correlation function is amplified, as expected.

\section{\label{sec:conclusion} Conclusions}
In this paper we investigated how the energy and momentum deposited by dijets in the quark-gluon plasma may affect the direct, elliptic, and triangular flow of low (and intermediate) $p_T$ hadrons in central Au+Au collisions at RHIC. The dijets are modeled as external sources in the energy-momentum conservation equations, which are solved event-by-event within boost invariant ideal hydrodynamics.

We can emphasize three aspects: (i) the effects of the dijets on the QGP seem to not be important in the region of low $p_T$ ($p_T < 1$ GeV). Even for the highest value of energy loss used in this paper, $dE/dl|_{0} =20$ GeV/fm, which corresponds to an average suppression of $83\%$ of the initial jet transverse energy, we found that dijets affect mainly the region of intermediate $p_T$ ($1 < p_T < 3$) GeV; (ii) for the same range in $p_T$, the correlation between the flow parameters $\left\{v_n,\Psi_n \right\}$ and the initial geometric parameters $\left\{\varepsilon_{m,n},\Phi_{m,n} \right\}$ is considerably reduced due to the fact that the anisotropic flow created by the dijets is not related to the transverse shape of the initial energy density distribution and (iii) the direct flow, $v_1$, created by the dijets can be clearly seen in the profile of the dihadron angular correlation function, especially if only events with jets are selected. We suggest to compare the dihadron angular correlation function obtained from two distinct ensembles, one with (at least one dijet) and another one without jets, in order to have a rough estimate of the magnitude of the coupling between the jets and the QGP.

We believe that the main features found here should also be present in more realistic simulations. We hope that our results motivate other studies of the effects of jets on the bulk anisotropic flow coefficients of the QGP using a more realistic model for the energy loss (instead of the simplified phenomenological model used here) and a more realistic hydrodynamical computation (full 3+1 dynamics with viscous effects and different models for the initial conditions).

R.~P.~G.~Andrade and J.~Noronha thank
Funda\c c\~ao de Amparo \`{a} Pesquisa do Estado de S\~{a}o Paulo
(FAPESP) and Conselho Nacional de Desenvolvimento Cient\'{\i}fico e Tecnol%
\'{o}gico (CNPq) for financial support. G. S. Denicol acknowledges
the support of a Banting fellowship provided by the Natural Sciences and Engineering Research Council of Canada. The authors thank F.~Navarra, A.~Dumitru, and S.~Jeon for comments.

\nocite{*}

\bibliography{draft}% Produces the bibliography via BibTeX.

\end{document}